\documentclass[10pt,draft]{dis03}
\usepackage{epsf,epsfig,amsmath,graphics}

\newcommand{\newc}{\newcommand}
\newc{\ra}{\rightarrow}
\def\xg{x_\gamma}
\newc{\ETJ}{E^{{\rm Jet}}_T}
 
\begin{document}

\title{Studying QCD in the Final State of High Energy Collisions}
%\thanks{This work is
%supported by the High-Energy Physics  Foundation}}

\author{J. M. Butterworth \\
Department of Physics and Astronomy, University College London,\\
and Hamburg University, Alexander von Humboldt Fellow\\
E-mail: J.Butterworth@ucl.ac.uk }

\maketitle

\begin{abstract}
\noindent 
I give an outline and some discussion of the results presented during
the hadronic final states parallel sessions of the meeting, and some
opinions about the current state of play and future directions in this
area.
\end{abstract}

\section{Event shapes and energy flows: 
Deep Inelastic Scattering {\it vs} e$^+$e$^-$} 
\label{sec:ee}

\begin{figure}[!thb]
\vspace*{12.0cm}
\begin{center}
\includegraphics{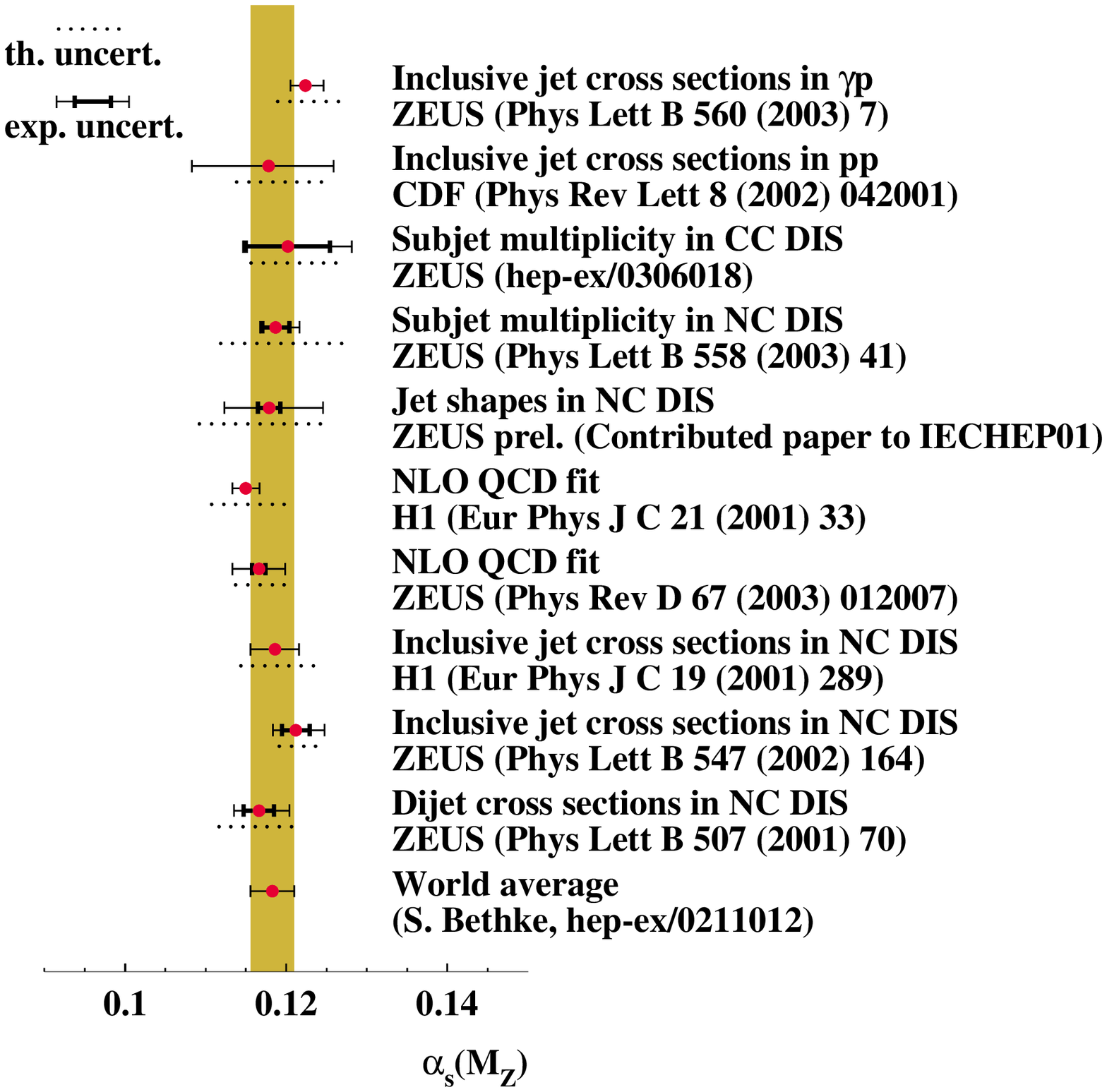}
\includegraphics{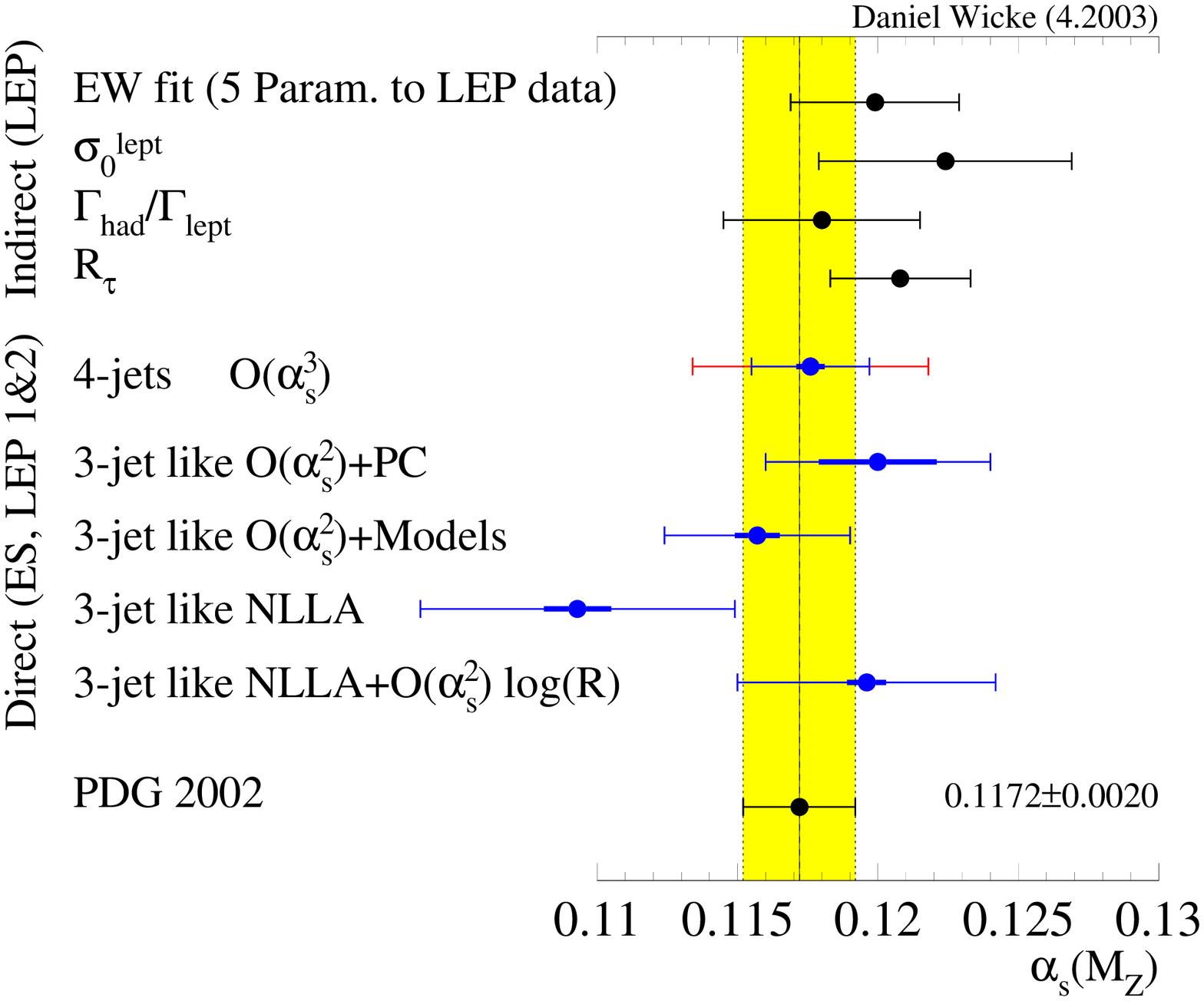}
\caption{ 
$\alpha_s$ measurements from hadron colliders and LEP.
\label{fig:as}}
\end{center}
\end{figure}

To go straight to the main result, the level of precision and
understanding achieved can be expressed in terms of measurements of
the strong coupling constant $\alpha_s$, the fundamental parameter of
QCD. A selection of such measurements~\cite{alphas,passon} is shown in
Fig.~\ref{fig:as}. The majority of the measurements shown are made
using jet cross sections and event shapes properties in the final
state of deep inelastic scattering (DIS) at events at HERA or $e^+e^-$
annihilation at LEP, though extractions from parton distributions, as
well as those from jet cross sections in $p\bar{p}$ and
photoproduction are also shown. More will be said of these in
Section~\ref{sec:hh}.

To make these measurements, many technical advances in the QCD
calculation of final states have been exploited. In particular, to
produce combined LEP results involved lots of dialogue between the
experiments and with theorists to obtain agreement on such issues as
choice of jet algorithm, treatment of correlations, power corrections
and treatment of hadronisation. Convergence on such issues is
necessary in order to gain reasonable confidence in the estimated
uncertainties on $\alpha_s$.  The HERA measurements are of similar
precision to the LEP results but have not yet been combined. Such a
combination of results certainly makes sense and is a challenge mainly
for the experimentalists. However, in common with the LEP results, it
is clear that theoretical uncertainties are at least as large, and
often larger than, the experimental errors. Understanding the
correlations between the theoretical uncertainties of different
measurements (for instance from jet rates, jet shapes, subjets and
fits to structure functions) is a challenge for phenomenologists as
well as experimentalists, and seems a necessary next step if such
measurements are to be fully exploited.

A big remaining issue is the fact that uncertainty from higher orders
can only be estimated from the scale dependence of NLO calculations.
Even if experiments agree on a common approach, it remains
arbitrary. In addition, perturbative calculations alone do not provide
a particularly good fit to the data at LEP~\cite{passon}, and
discrepancies are also seen in the H1 data~\cite{specka} on jet
production at low $Q^2$.  There is still an active debate on how the
effects of higher orders and non-perturbative physics are best
estimated~\cite{passon}. In particular, treating the scale as a free
parameter leads to a good description of the LEP data with fixed-order
perturbative calculations alone. However, the physical meaning of such
a procedure and the uncertainties it involves remain unclear.  An
obvious though arduous improvement would be NNLO calculations of some
key processes. Other possibilities for improved accuracy are the
inclusion of power corrections, and the established technique of
resumming key contributions to the cross section to all orders in
$\alpha_s$.

In the latter area, progress following the recent theoretical
discovery of non-global logarithms \cite{ngl} was
reported. Significant large logarithms can occur for observables made 
in a limited region of phase space. Since detector acceptance usually
limits measurement to sub-regions of phase space, this is a rather
common type of variable!  Amongst those expected to be affected are
single jet profiles, jet substructure, cuts involving rapidity gaps or
just rapidity cuts driven by acceptance, prompt photon isolation cuts
based on energy within a cone, interjet energy flows and more.  The
discovery of these logarithms opens a new phenomenlogical avenue in
QCD which is now being actively studied. The question as to whether
the improved understanding embodied in this work will lead to more
accurate predictions remains open.

To gain the full benefit of new advances in our understanding of QCD
without prohibitive cost in theoretical working hours, technical
improvements and automation become important. Furthermore, in my
opinion, the process of generalizing a technique to the extent that it
can be automated is one of the more convincing demonstrations that a
real understanding has indeed been achieved. An example of such was
given with the CAESAR resummation program, and some first results were
reported~\cite{banfi}. This program also applies to event shapes in
hadron-hadron collisions. Resummation of large logarithms in event
shape variables is needed to make accurate predictions and to describe
the data. However, though the theory is `known', each variable
requires a new calculation (on average one paper per variable). The
systematic algorithmic approach adopted in CAESAR means a computer can
take over. The user provides a routine which calculates the variable,
CAESAR makes sure it can be resummed with current theoretical
understanding. The result is the equivalent of an analytic
calculation, with which one can study scale dependence, apply
hadronisation corrections and so on. In addition the result is free of
subleading logs and can be matched to NLO matrix elements in
principle.

Other technical improvements reported include a master formula for
NNLO soft and virtual QCD corrections, calculations of two-loop and
n-loop eikonal vertex corrections \cite{kidonakis}, and resummation in DIS
and Drell-Yan~\cite{magnea}. There was also a discussion of
rescattering effects on DIS parton distributions~\cite{hoyer} and
universality-breaking effects in diffractive DIS and Drell-Yan
production~\cite{peigne}. 

\section{Jets and energy flows: 
$p\bar{p}$ {\it vs} $\gamma p$ {\it vs} $\gamma\gamma$}
\label{sec:hh}

Lepton beams provide a spectrum of low-virtuality photons.  These
photons can fluctuate into virtual $q\bar{q}$ pairs and acquire a
hadron-like structure, allowing lepton-lepton and lepton-hadron
colliders to mimic hadron-hadron colliders.  There is much to be
learned from comparisons between jet or energy flow measurements (as
well as heavy flavour production, see~\cite{sefkowetc}) in $p\bar{p}$
(Tevatron), $\gamma p$ (HERA) and $\gamma\gamma$ (LEP).

As some compensation for their lower centre-of-mass energy, LEP and
HERA can also make cleaner measurements with a pointlike photon. Going
to high photon virtuality (in Deep Inelastic Scattering) or to high
$\xg$ (direct photoproduction) allows the photon's hadron-like
structure to be effectively ``turned off''. In such cases effects of,
for example, the underlying event (see below) can be greatly altered or
removed.

\subsection{Tevatron Data}
\label{sec:pp}

As well as a review of some Run I data~\cite{skatchkov,lami}, some
first results on jets, dijets and shapes where shown from Tevatron Run
II~\cite{bhatti,odell}.  The improvements include the higher beam
energy (which gives a significant increase in the cross section at
very high transverse energy) and better triggers. The calorimeter
energy scale is currently known to around 5\%, though this is expected
to improve in the near future.  Measurements of fully corrected cross
sections - i.e. with detector effects removed - were presented and
offer the hope of precision jet measurements in the near future. The
measurements of inclusive jets now extend up to 550~GeV.  There is no
sign of any excess with respect to NLO QCD at high $\ETJ$, and
comparison with Run I data \cite{tevinc} shows good agreement between
the data sets.  Dijet mass distributions have also been measured using
the new data, and compared to NLO QCD for masses up to 1364 GeV, again
showing good agreement.

Some early results on jet shapes were shown. The results were not
corrected for detector effects, but HERWIG and PYTHIA, both containing
leading-logarithmic parton showers, were passed through a detector
simulation for comparison, and describe the data quite well. There is
a noticeable broadening of jets at high rapidities.

\subsection{HERA $\gamma p$ Data}

Jets in photoproduction have been measured up to $\ETJ=90$~GeV at HERA
\cite{specka,incz} with dijet pairs up to masses of 140~GeV
\cite{dij}. The effects of hadronisation and underlying events are
important below about around 15~GeV, but not above. The calorimeter
energy scale is known to around 1\% for these energies, and the
measurements are thus rather precise. An example of the precision
achieved here is given by measurements of scaling violations in
photoproduction~\cite{incz}, made by exploiting the variable-energy
photon beam at HERA, which give a rather accurate determination of
$\alpha_s$ (see Fig.~\ref{fig:as}). Inclusive and dijet measurements
are sensitive to the PDF of the photon and the proton, particularly
the gluon in the proton at high $x$. They should be included in future
global fits of photon and proton PDFs. This is an issue of some
importance for understanding QCD backgrounds both at a future linear
$e^+e^-$ collider and at the LHC.

Prompt photons have also been measured in photoproduction and DIS at
HERA~\cite{lemrani} - new results from H1 and ZEUS (respectively) were
presented. There is good agreement between the collaborations and with
NLO pQCD for photoproduction. The NLO calculation also describes the
first DIS measurement reasonably well, at least in
normalisation. However, there are differences in some trends, and the
LO Monte Carlos completely fail (in different ways!) to describe this
process.  Since prompt photons give, in principle at least, more
direct access to the hard scattering process, they are a good testing
ground for QCD and for determination of PDFs, as well as being a
calibration tool at hadron colliders. The isolation requirement leads
to sensitivity to QCD final state effects which is different from that
seen in jet cross sections - underlying events reduce the cross
section, rather than increasing it. 

\subsection{LEP $\gamma\gamma$ Data}

Extensive and precise OPAL results on dijet production in
$\gamma\gamma$ collisions were shown \cite{krueger}. Jet shapes and
jet cross sections have been measured and events have been separated
into regions dominated by collisons between pointlike photons,
hadronic photons or one of each. There is in general good agreement
with NLO QCD calculations, though some discrepancies are seen at low
$\ETJ$ when both photons are hadronic (see below for some discussion
of this). Results on inclusive hadron production and jet cross
sections in two-photon events from L3 were also
presented~\cite{l3}. These data are generally consistent with OPAL
(where the kinematic regions overlap), and with NLO QCD except at the
highest transverse energies, where the inclusive hadron and jet data
both lie above the calculation. OPAL do not measure to such high
transverse energies. This is the most striking discrepancy between the
data and NLO QCD reported during the sessions, and clearly requires an
explanation.

\vspace{0.5cm}

\noindent
In summary of this section - in most cases NLO QCD calculations for
the important observables are there, and generally describe the
data. However, there is a catch in that in many, even most, cases the
data are more accurate than the available predictions. This is a
similar situation to that discussed in Section~\ref{sec:ee}. This is
not necessarily an indication of a lack of understanding of QCD, since
many improvements in the calculations are possible in principle. Hence
the importance of practical improvements such as those discussed in
Section~\ref{sec:ee} is worth reiterating.

As well as higher order calculations and resummations, it is also
important that non- and semi-perturbative effects can be isolated. In
some cases it is beneficial to avoid these effects, but they remain
important and interesting in their own right in several
areas. Hadron-hadron collisions at the LHC are the near future of high
energy physics, and it is becoming clear that quantitative and precise
information on hadronic final states from Tevatron, HERA and LEP will
be essential to fully exploit the opportunities of this new
machine. An example of the kind of studies which are now beginning in
several places was given in this session~\cite{skatchkov}.

In this light, there are some contrasts between the three major
colliders discussed in the meeting. The Tevatron has the advantage of
higher centre-of-mass energy and actual proton beams. The QCD analyses
from Run I were of course limited by the understanding at the time and
used methods (such as choice of jet algorithm and underlying event
corrections) which make quantitative comparisons with theory
difficult. Unfortunately, despite the advances in phenomenological
understanding between Run I and Run II, the (infrared-unsafe) cone
algorithm still appears to be at least the default choice, and the
collaborations are sometimes still correcting for energy outside jet
(``underlying event'') which is actually part of current QCD
calculations. These issues are not important for the present data, but
if the hoped-for understanding of the detectors is reached, it will
become important to use more robust and model-independent techniques
to control uncertainties in the comparison with theory. There is a lot
of potential here, and there are people on the experiments battling to
realise it. At HERA there are mature analyses now being
published. High precision has been achieved (even in photoproduction!)
and there has been overall a fairly good take-up of new
phenomenological developments. New data from HERA II will extend the
precision region up in transverse energy closer to the kinematic
limit.  The data from LEP2 which are still being published show in
general statistically limited but elegant and exciting results. The
effort available to analyse the data is inevitably winding down, and
the question is, has everything that is needed really been done?  At
least the discrepancy between L3 data and NLO QCD should be
understood, and measurements at the highest transverse energies from
other LEP experiments would help.

An area of common concern in these studies is the effect of underlying
events and minimum bias data on final state observables such as jet
cross sections, jet shapes and energy flows between jets. This is even
an issue at lepton colliders - for example, in the subset of OPAL
$\gamma\gamma$ data \cite{krueger}, where both photons are hadronic,
NLO QCD lies below the data, whereas elsewhere it is in good
agreement. This is exactly what one would expect of a significant
contribution from secondary scattering between the photon
remnants. Similar effects are observed in Tevatron and HERA data, and
indeed direct evidence for multiple hard scattering has been published
by CDF~\cite{cdfmi}.  QCD based models for this physics exist, and
have been implemented in Monte Carlo programs~\cite{mcs}, but they
need to be fairly complex to describe the correlations, fluctuations
and energy dependencies seen in the data. These models are currently
being tested and tuned using Tevatron, HERA, LEP and older
hadron-hadron data~\cite{tuners}, and for example, the latest CDF
minimum-bias data~\cite{lami,field} have only been successfully
described by simulations including such physics.  A reasonable
understanding of such effects is a necessary input for detector
development and precision measurements at current machines, LHC and
FLC.

\section{Fragmentation, resonances and  non-perturbative effects}

There were presentations on scalar meson resonance production from
HERMES~\cite{garutti}, H1~\cite{kropivnitskaya} and ZEUS~\cite{raval}.
Scalar mesons can be glueball candidates, and since there is evidence
for too many of them to fit into the expected multiplets, studying
their production cross sections in different environments is of
particular interest. HERMES and H1 observe the $f_0(980)$ in the
$\pi\pi$ decay channel. H1 also see the $f_2$, and ZEUS, looking at
$K^0_sK^0_s$ decays see the $f_2(1525)$ and what is probably the
$f_0(1710)$. This latter, which is a favoured glueball candidate, is
seen in the target region of the Breit frame in DIS.  ZEUS have also
shown new results on strangeness production in DIS which show signs of
an interesting dependence on kinematic region, in particular in the
target region of the Breit frame~\cite{raval}.

Moving to particle production from nuclear targets, a model of
quenching of hadron momentum spectra in DIS on nuclear targets was
presented~\cite{arleo} . In this model it is assumed that all
quenching is due to rescattering of partons in nuclei before
hadronisation rather than after. This means that such effects can
be treated in partonic calculations, and the model makes some
definite predictions, one of which is saturation of these effects
for large nuclei (since above some radius the rescattering length is
determined by the length of time to hadronise, not by nuclear
size). Another is that and $K^+$ and $K^-$ production should be
equally suppressed. These variables are proposed as a very sensitive
way of distinguishing between this model and others where
quenching takes place after hadronisation.

There were also presentations on Nuclear attenuation in semi-inclusive
electroproduction of hadrons at HERMES~\cite{elkebian} and on searches
for QCD instantons at ZEUS~\cite{olkiewicz}.

Two speakers presented new results on Bose-Einstein Correlations, as
well as a compliation of a wide variety of older LEP
results~\cite{olkiewicz,boutemeur}. These correlations are fascinating
and fundamental quantum-mechnical effect which give a rather direct
probe of the hadron production region in high energy collisions. This
of course is a poorly understood and important area in QCD, being
where confinement takes place.  The new OPAL (and older L3) analysis
elegantly avoids coulomb effects by using neutral pion pairs. Of
particular interest is whether the size and shape of the hadron
production region depends strongly on the hard scattering process. The
ZEUS~\cite{olkiewicz} data, using inclusive hadrons in DIS, indicate
that there is no such dependence, since the effect is shown to be
constant over a very wide range of $Q^2$.

Results on colour reconnection show how non-perturbative QCD may
affect the $W$ mass measurement at LEP New studies in $e^+e^- \ra
W^+W^-$ at $\sqrt{s} = 189 \ra 208$~GeV using particle flow were
presented \cite{leibenguth}. Several models can now be ruled out, and
no definite effect seen in $Z$ or $WW$.  The end effect on the $W$
mass is $22 \pm 43$~MeV.

\section{Unintegrated parton distributions and low $x$}
 
Low-$x$ and diffractive physics were covered in detail in another
working group \cite{diff}.  However, some areas where the hadronic
final state can be used to investigate this physics were discussed in
this session.

A general motivation is the search for a connection between the
behaviour of total cross sections at high energies (high $s$) and
low-$x$ partonic physics.  As $s/t$ becomes large, $x$ becomes small
and large rapidity intervals are available in the hadronic final
state. It is interesting to study whether calculations which include
to all orders (resum) terms containing large logarithms of $1/x$, or
(equivalently) large rapidities, describe the data better than those
which only resum large logarithms in the hard scale $Q^2$. A feature
of the former calculations is that they break strong ordering in the
scale, and thus parton distributions should explicitly include
significant incoming parton virtuality, or transverse momentum, rather
than their integratal. Cross sections at high $s$ have in the past
been described using Regge phenomenology, and so studying low-$x$
physics may determine whether or not Regge phenomenology is an
emergent property of QCD.
 
One area where such physics may appear is in high rapidity jets in
DIS. New results from H1~\cite{goerlich} and ZEUS~\cite{lammers} were
presented. The ZEUS results show an interesting excess in the
inclusive jet cross section at high jet rapidities. When further cuts
are applied to ensure that Bjorken $x$ is small for the whole event,
such that there is indeed a large evolution in $x$ between the photon
vertex and the forward jet, the uncertainties in the NLO QCD
calculation increase (as indeed might be expected if the series is not
converging so well due to large logarithms in $x$), and though the
data still lie above the calculation, they are within these
uncertainties.

The status of the CASCADE Monte Carlo, which uses the CCFM equation
and unintegrated parton distributions was presented~\cite{cascade}. New
fits are now available, including a better treatment of soft regions
of the cascade using cut offs, and non-leading contributions. The new
fits give better agreement with new H1~\cite{goerlich} forward jet
data.

New results on rapidity gaps and energy flows between jets in
photoproduction were also shown by ZEUS~\cite{sutton}. These extend
previous measurements~\cite{gaps} of this process at HERA and confirm
evidence for hard colour singlet exchange in these collisions.  This
is an exciting process since as with the forward jets in DIS, the
momentum exchange across the rapidity interval (a gap, in this case)
is much greater than $\Lambda_{\rm QCD}$ and so perturbative QCD
should apply; but at the same time, $s/t$ is large, and the process is
in some sense diffractive.

New H1 data on dijet production at low Bjorken-$x$ in DIS
\cite{poeschl} show an increasing decorrelation of dijet pairs as $x$
decreases, as is qualitatively expected if low-$x$ terms are becoming
important. Three-jet LO calculations do not describe the data, though
NLO three jet calculations~\cite{nagy} do much
better~\cite{h1eps}. Nevertheless, there is still a very interesting
excess in the data at the lowest $x$ and $Q^2$.

There was a presentation showing that use of Regge theory to determine
the starting parton distributions and DGLAP evolution to describe
their $Q^2$ dependence was shown to describe the latest inclusive DIS
data rather well~\cite{soyez}. Finally there was a presentaion on the
effect of unintegrated gluon distributions on inclusive particle
production in hadronic collisions at SPS~\cite{szczurek}. In the
latter case it seems that the effects could be significant. Such
studies again motivate a broad approach to model comparisons across
hadronic final states in different processes and at different
energies.

\section{Summary}
Making quantitative QCD studies in final states is technically
challenging for experiment and theory. Lots of excellent new
measurements have been presented during the meeting, along with
promises of more to come. There have also been some significant
advances in theory and phenomenology.  Next-to-leading order QCD is in
general needed to provide a satisfactory understanding of current
data. Where such calculations exist, perturbative QCD is doing well,
though within large theoretical uncertainties. Non- or semi-
perturbative effects are being studied quantitatively and there are
interesting models on the market.  Incremental but real progress
continues to be made in this critically important area of high energy
physics.

\section*{Acknowledgements}
My thanks to the organisers, my co-convener Yuri Dokshitzer, and the
working group participants for a very friendly and interesting
conference. Thanks also to C. Glasman, M. Sutton and R. Yoshida for
comments and help with this write up.

\end{document}